\input phyzzx.tex
\tolerance=1000
\voffset=-0.0cm
\hoffset=0.7cm
\sequentialequations

\def\t1{{\tilde 1}}

\def\t{\theta}


\REF{\ABH}{M. Cadoni and S. Mingemi, Phys. Rev. {\bf D51} (1995) 4319, [arXiv:hep-th/9410041].}
\REF{\MAL}{J. Maldacena, D. Stanford and Z. Yang, PTEP 2016 (2016) no.12, 12C104, [arXiv:1606.01857].}
\REF{\ADS}{J. Maldacena, Adv. Theor. Math. Phys. {\bf 2} (1998) 231, [arXiv:hep-th/9711200]; S. Gubser, I. Klebanov and A. Polyakov, Phys. Lett. {\bf B428} (1998) 105, [arXiv:hep-th/9802109]; E. Witten, Adv. Theor. Math. Phys. {\bf 2} (1998) 253, [arXiv:hep-th/9802150].}
\REF{\PGB}{G. J. Turiaci and H. L. Verlinde, JHEP {\bf 1612} (2016) 110 [arXiv:1603.03020]; E. Halyo,  [arXiv:1907.08149].}
\REF{\CHI}{A. Strominger, JHEP {\bf 9901} (1999) 007, [arXiv:hep-th/9809027].} 
\REF{\CQM}{D. Gaiotto, A. Strominger and X. Yin, JHEP {\bf 0511} (2005) 017, [arXiv:hep-th/0412322].}
\REF{\VJY}{V. Balasubramanian, J. de Boer,  M. M. Sheikh--Jabbari and J. Simon, JHEP {\bf 1002} (2010) 017, [arXiv:0906.3272].}
\REF{\JT}{R. Jackiw, Nucl. Phys. {\bf B252} (1985) 343; C. Teitelboim, Phys. Lett. {\bf B126} (1983) 41.}
\REF{\GB}{G. Gibbons and S. H. Hawking, Phys. Rev. {\bf D15} (1977) 2752.}
t\REF{\POL}{A. Almheiri and J. Polchinski, JHEP {\bf 1511} (2015) 014, [arXiv:1402.6334].}
\REF{\MET}{M. Cadoni and S. Mingemi, Phys. Rev. {\bf D59} (1999) 081501, [arXiv:hep-th/9810251]; Nucl. Phys. {\bf B557} (1999) 165, [arXiv:hep-th/9902040].}
\REF{\VER}{J. Engelsoy, T. Mertens and H. Verlinde, JHEP {\bf 1607} (2016) 139, [arXiv:1606.03438].}
\REF{\BTZ}{M. Banados, C. Tietelboim and J. Zanelli, Phys. Rev. Lett. {\bf 69} (1992) 1849.}
\REF{\WIT}{E. Witten, Adv. Theor. Math. Phys. {\bf 2} (1998) 505, [arXiv: hep-th/9803131].}
\REF{\EIN}{G. Jorjadze, Liouville Field Theory and 2D Gravity, 1998.}
\REF{\CARL}{S. Carlip, Phys. Rev. Lett. {\bf 82} (1999) 2828, [arXiv:hep-th:9812013]; Class. Quant. Grav. {\bf 16} (1999) 3327,
[arXiv:gr-qc/9906126].}
\REF{\HOR}{E. Halyo, [arXiv:1502.01970]; [arXiv:1506.05016].}
\REF{\CON}{E. Halyo, [arXiv:1503.03137].}
\REF{\RIN}{E. Halyo, [arXv:1809.10672].}
\REF{\LIO}{see for example H. Erbin, Notes on 2D Quantum Gravity and Liouville Theory, 2015.}
\REF{\SOL}{S. Solodukhin, Phys. Lett. {\bf B454} (1999) 213, [arXiv:hep-th/9812056].}
\REF{\EDI}{E. Halyo, [arXiv:1606.00792].}
\REF{\STR}{A. Strominger, JHEP {\bf 9802} (1998) 009, [arXiv:hep-th:9712251].}
\REF{\NEW}{E. Halyo, in preparation.}

\singlespace
\pagenumber=0
\normalspace
\medskip
\bigskip
\titlestyle{\bf{Are Near $AdS_2$ Spacetimes Dual to $AdS_2$ Black Holes?}}
\smallskip
\author{ Edi Halyo{\footnote*{email: halyo@stanford.edu}}}
\smallskip
\centerline {Department of Physics} 
\centerline{Stanford University} 
\centerline {Stanford, CA 94305}
\smallskip
\vskip 2 cm
\titlestyle{\bf Abstract}

We claim that there is a one to one correspondence between $NAdS_2$ spacetimes and dilatonic $AdS_2$ black holes. We show that these have the same thermodynamics and can be described by the same states of 1D CFTs with the same central charge. In addition,
we show that these are actually the same set of solutions since a field redefinition in the Schwarzian theory transforms its equation of motion into the time independent part of the bulk Einstein equation. The deformations of the $AdS_2$ boundary as a function of the boundary time are exactly the conformal factor of the $AdS_2$ black hole metric as a function of the radial bulk coordinate.

\singlespace
\vskip 0.5cm
\endpage
\normalspace

\centerline{\bf 1. Introduction}
\medskip

Two dimensional dilatonic gravity with a negative cosmological constant is an interesting toy model of quantum gravity. 
This is mainly due to the existence of dilatonic $AdS_2$ black hole solutions with nonzero temperature and entropy[\ABH]. Recently 
new solutions of the same theory, including the Gibbons--Hawking boundary term, were discovered. These are near $AdS_2$ ($NAdS_2$) spacetimes which basically describe the deformations the $AdS_2$ boundary[\MAL]. Just like the black holes, $NAdS_2$ spacetimes are thermal with nonvanishing temperature and entropy.
Naively, these two sets of solutions seem different since they have different metrics with different boundaries. In this paper, we show that there is a one to one correspondence between dilatonic $AdS_2$ black holes and $NAdS_2$ spacetimes. Moreover, 
we show that, after a field redefinition in the 1D boundary theory, the deformations of the $AdS_2$ boundary (as a function of boundary time) described by $NAdS_2$ spacetimes are exactly the same as the conformal factor (as a function of the bulk radial coordinate) of the $AdS_2$ black hole metric. Thus both sets of solutions are defined by the same function.

The AdS/CFT duality[\ADS] requires that bulk $AdS_2$ solutions such as black holes be dual to states in the 1D boundary CFT. In this paper, we claim that $NAdS_2$ spacetimes, which describe deformations of the boundary, are these boundary states.  
As stated above, both $AdS_2$ black holes and $NAdS_2$ spacetimes are thermal solutions, i.e. they have nonzero temperature and entropy. It is easy to show that, with a proper choice of the asymptotic boundary value of the dilaton, both sets of solutions have the same equation of state, $S(M)$ and therefore the same thermodynamics. This implies that there is a one to one correspondence between these two sets of solutions.

Further support for this correspondence comes from the fact that both $AdS_2$ black holes and $NAdS_2$ spacetimes can be described as states with the same conformal weights in 1D CFTs with the same central charge. 
Even though the details of the 1D boundary CFT dual to $AdS_2$ is not known, its central charge can be deduced from the asymptotic symmetry of the metric[\ABH]. 
The conformal weight of a state that corresponds to a black hole is fixed by its mass. Using these, the Cardy formula reproduces the correct black hole entropy. On the other hand,
$NAdS_2$ spacetimes are described by the Schwarzian action[\MAL]. In refs. [\PGB], it was shown that a 1D CFT can be described in terms of the pseudo Goldstone bosons(PGBs) of conformal symmetry with the same action. In this case, the central charge of the CFT and the conformal weight of the state are fixed by the coefficient of the Schwarzian action and the mass of the $NAdS_2$ spacetime respectively. The central charges and conformal weights of both sets of solutions match as required from a one to one correspondence. 

Finally, we show that $AdS_2$ black holes and $NAdS_2$ spacetimes  are not only in a one to one correspondence with each other but are actually the same set of solutions in the following sense. In the conformal gauge, the $AdS_2$ black hole metric is defined by a single function, i.e. the conformal factor which is only a function of the radial bulk coordinate. On the other hand, $NAdS_2$ spacetimes, which are solutions of the Schwarzian boundary theory, are also defined by a single function of the boundary time.
With a proper field redefinition and identification of the boundary time with the radial bulk coordinate, the boundary equation of motion becomes exactly the time independent part of the 2D Einstein equation with a negative cosmological constant. $NAdS_2$ spacetimes are solutions of the former whereas $AdS_2$ black holes are solutions of the latter. 
Thus, the same procedure takes each deformation of the $AdS_2$ boundary described by an $NAdS_2$ spacetime (with a $T$) into the conformal factor of a dilatonic $AdS_2$ black hole (with the same $T$). We find that both sets of solutions are described by the same function.

The AdS/CFT duality implies that the $AdS_2$ black holes are dual to states in the 1D boundary CFT. The nature of this boundary theory is not well known; possible candidates include a 2D chiral CFT[\CHI], conformal QM[\CQM] and DCLQ limit of a 2D CFT[\VJY]. The one to one correspondence between $AdS_2$ black holes and $NAdS_2$ spacetimes, which are simply deformations of the $AdS_2$ boundary, leads us to conjecture that the latter are holographically dual to the former. If this is correct, then the 1D boundary CFT is described by the Schwarzian action. The fact that the AdS/CFT dual bulk and boundary states, i.e. $AdS_2$ black holes and $NAdS_2$ spacetimes are actually the same set of solutions, in the sense described above, is unique to 2D.

This paper is organized as follows. In the next section, we briefly review 2D dilatonic $AdS_2$ black holes and $NAdS_2$ spacetimes.
In section 3, we show that these two sets of solutions have the same thermodynamics. In section 4, we show that both the $AdS_2$ black holes and $NAdS_2$ spacetimes can be described as states with the same conformal weight in 1D CFTs with the same central charge. In section 5, we show that $AdS_2$ black holes and $NADS_2$ spacetimes are actually the same set of solutions. Section 6 contains a discussion of our results and our conclusions.

\bigskip
\centerline{\bf 2. $AdS_2$ Black Holes and $NAdS_2$ Spacetimes}
\medskip

In this section, we review dilatonic $AdS_2$ black holes and $NAdS_2$ spacetimes which we explore in the following. Both arise from the 2D gravitational action with a negative cosmological constant including the Gibbons--Hawking boundary term[\GB]. The black hole solutions are obtained from the bulk action without the boundary term. On the other hand, $NAdS_2$ spacetimes are deformations of the boundary that naturally arise from the boundary term. These have different metrics and boundaries and therefore, naively, appear to be two different sets of solutions. However, as we will show in this paper, they are in a one to one correspondence and in fact are defined by the same function.

We begin with the dilatonic $AdS_2$ gravity with the (Euclidean) action
$$I=-{1 \over 2} \int d^2x \sqrt{g} \phi \left(R+{2 \over \ell^2}\right) \quad, \eqno(1)$$
where $\phi$ is the dilaton field and the cosmological constant is given by $\Lambda=-2/\ell^2$. 
In eq. (1) we neglected the boundary term since we are interested in the bulk solutions.
This is the Jackiw--Teitelboim (JT) theory[\JT] and has dilatonic 
black holes with the metric[\MET]
$$ds^2= \left({r^2 \over \ell^2}-{{2M\ell} \over \phi_0}\right)dt^2+\left({r^2 \over \ell^2}-{{2M\ell} \over \phi_0}\right)^{-1}dr^2
\quad, \eqno(2)$$
and the linear dilaton profile $\phi=\phi_0 r/\ell$. Here $M$ is the black hole mass and $\phi_0$ fixes the two--dimensional Newton constant by $\phi_0=1/8 \pi G_2$. The black hole horizon is at $r_h=(2M \ell^3/\phi_0)^{1/2}$. 
We can also write the black hole metric in the conformal gauge as
$$ds^2=e^{\rho(x_1)}(dx_0^2+dx_1^2) \qquad \qquad \rho(x_1)= {{\mu^2 \ell^2} \over {sinh^2 \mu x_1}} \qquad, \eqno(3)$$
where $\rho(x_1)$ is the conformal factor and $\mu=\sqrt{4 \pi G_2 M/\ell}=r_h/2 \ell^2$. In these coordinates, the dilaton solution becomes $\phi(x_1)=\mu \ell coth(\mu x_1)$.

There are two special cases of the black holes in eq. (2). First, in the $M=0$ limit, the black hole reduces to the Poincare patch of $AdS_2$. Second, the black hole with $M=-16 \pi G_2/ \ell$ corresponds to global $AdS_2$ spacetime. Solutions with $-16 \pi G_2/\ell<M<0$ are not physical since they lead to naked singularities.

In order to describe $NAdS_2$ spacetimes, which are deformations of the $AdS_2$ boundary, we need to add the Gibbons--Hawking boundary term[\GB] to the action in eq. (1). The action can then be written as $I=I_{ext}+I_{nonext}$ where the extremal and nonextremal parts are[\MAL,\POL] 
$$I_{ext}=-{\phi_0 \over 2} \int d^2x \sqrt{g}R+{2 \over {\ell^2}} \int_{bndy} du K \quad, \eqno(4)$$
and 
$$I_{nonext}=-{1 \over 2} \int d^2x \sqrt{g} \phi (R+{2 \over {\ell^2}})+2 \int_{bndy} du \phi_{b}K \quad. \eqno(5)$$
The one--dimensional boundary of $AdS_2$ is parametrized by the Euclidean boundary time $u$. Here $\phi_b$ is the value of
the dilaton at the boundary.
The extremal action in eq. (4) is topological (equal to $4 \pi$) and not relevant for our purposes.

Unlike the black hole case above, we can now impose
the dilaton equation of motion, i.e. $R=-2/\ell^2$ which fixes spacetime to be $AdS_2$.
Therefore, the first term in eq. (5) vanishes and we are left with the boundary action.
In ref. [\MAL], 
it was shown that, when the bulk coordinates of the Poincare patch of $AdS_2$, $t(u),z(u)$ are considered to be functions of the boundary time $u$, the boundary action in eq. (5) can be written as
$$I_{bndy}= - \int_0^{\beta} du \phi_r(u) Sch(t,u) \quad, \eqno(6)$$
where the Schwarzian derivative is given by
$$Sch(t,u)=\left({t^{\prime \prime} \over {t^{\prime}}} \right)^{\prime}- {1 \over 2} {t^{\prime \prime 2} \over {t^{{\prime}2}}} \quad, \eqno(7)$$
and prime denotes a derivative with respect to $u$. $\phi_r(u)=\epsilon \phi_b(u)$ is the normalized (finite) value of the dilaton at the boundary, $\epsilon \to 0$.
The Schwarzian action describes the physics of $t(u)$ which are the PGBs of the conformal symmetry which is spontaneously broken down to $SL(2,R)$. Thus, it is $SL(2,R)$ invariant and its solutions that are related by $SL(2,R)$ transformations to each other are equivalent. We will use this freedom to write our $NAdS_2$ solutions
in forms that are useful for our purposes. Each solution $t(u)$ (up to $SL(2,R)$ transformations)
corresponds to a different deformation of the $AdS_2$ boundary.
With the reparametrization $t(u)=cot(\tau/2)$ the action for $\tau$ becomes
$$I({\tau})=-\phi_r \int_0^{\beta} du \left ({1 \over 2}\tau^{\prime 2}+ Sch(\tau,u) \right) \quad. \eqno(8)$$
Above we used the fact that $\phi_r(u)$ can be set to any constant value ${\bar \phi}_r$ by reparametrizing the boundary time $u$ as $d{\bar u}={\bar \phi}_r du/\phi_r(u)$. With an abuse of notation we discard all bars from $u$ and $\phi$ in eq. (8) and the following.

The solution to the equation of motion for $\tau(u)$[\MAL], 
$$\tau^{\prime \prime}+{(Sch(\tau,u))^{\prime} \over {\tau^{\prime}}}=0 \quad, \eqno(9)$$
is $\tau(u)= \alpha u$ (up to an irrelevant additive constant) where $\alpha$ is a constant. The $AdS_2$ boundary (in Poincare coordinates) which is at $z=0$ before the deformation is given by $z(u)=t^{\prime}(u)$ in the $NAdS_2$ spacetime.
If we set $\alpha=2 \pi T$ and thus $t=cot(\pi T u)$ we obtain the thermal solution with nonzero entropy. As we will show below, this corresponds to the $AdS_2$ black hole solution.
This is the same solution given in ref. [\MAL], i.e. $t=tan(\pi T u)$, and is related to it by the $SL(2,R)$ transformation 
$t \to -1/t$ and $t \to -t$ which are symmetries of the Schwarzian action. As a result, both $t=cot(\pi T u)$ and 
$t=tan(\pi T u)$ have the same mass and entropy (see eqs. (12) and (13) below) as expected.
The two other $NAdS_2$ solutions we will be interested in are $t=1/u$ and $t=cot(-i u /\ell)$ which, as we will show below, correspond to the Poincare patch and global $AdS_2$ respectively. It is not clear whether these solutions result in unitary theories on the boundary. Nevertheless, we will consider them since their masses and entropies match their respective bulk solutions.


\bigskip
\centerline{\bf 3. Thermodynamics of $AdS_2$ Black Holes and $NAdS_2$ Spacetimes} 
\medskip

Above, we reviewed two sets of solutions of the action eq. (5) namely dilatonic $AdS_2$ black holes and $NAdS_2$ spacetimes.  Naively, these look completely different since the metrics and asymptotic behavior of the solutions are different.
In this section, we show that the thermodynamics of $AdS_2$ black holes and $NAdS_2$ spacetimes are the same. This is the first hint that these two sets of solutions are, at least, in a one to one correspondence. In the context of the AdS/CFT correspondence,
this should not come as a surprise if we consider $NAdS_2$ spacetimes as the boundary states which are dual to the bulk $AdS_2$ black holes.
 
The mass, temperature and entropy of dilatonic $AdS_2$ black holes are given by[\ABH]
$$M_{BH}={r_h^2 \over {16 \pi G_2 \ell^3}} \qquad T_{BH}={r_h \over {2 \pi \ell^2}} \qquad S_{BH}={r_h \over {4G_2 \ell}} \quad , \eqno(10)$$
or in terms of $T_{BH}$ we have
$$S_{BH}={{\pi \ell T_{BH}} \over {2G_2}} \qquad M_{BH}={{\pi \ell T_{BH}^2} \over {4G_2}} \quad . \eqno(11)$$
The mass of the $NAdS_2$ solution with $t=cot (\pi T u)$ is[\MAL,\POL]
$$M=\phi_b Sch(\tau,u)=2 \pi^2 \phi_b T^2 \quad. \eqno(12)$$
The entropy of the the $NAdS_2$ spacetime can be obtained by setting $logZ=-I({\tau})=2 \pi^2 \phi_r/\beta$ and using the standard formula[\MAL]
$$S=(1-\beta \partial_{\beta})logZ=4 \pi^2 \phi_r T \quad . \eqno(13)$$
We see that the mass and entropy of the $NAdS_2$ solution in eqs, (12) and (13) agree with those for $AdS_2$ black holes in eq. 
(11) if we set the normalized boundary value of the dilaton as
$\phi_r=\ell/8 \pi G_2$.
This shows that both $AdS_2$ black holes and $NAdS_2$ spacetimes have the same thermodynamics. As noted above, this is quite surprising. 
From the thermodynamical point of view, each $AdS_2$ black hole with mass $M$ is identical to an $NAdS_2$ solution with 
$t(u)=cot(\pi T u)$ where $T$ is the temperature of the black hole. This establishes the one to one correspondence between the two sets of solutions.
 
We now consider the two special black holes, namely black holes with $M=0$ and $M=-16 \pi G_2/\ell$
that correspond to the Poincare patch of $AdS_2$ and global $AdS_2$ respectively. In terms of $NAdS_2$, the Poincare patch of $AdS_2$ 
corresponds to the solution
with $t=1/u$. This has a vanishing Schwarzian and therefore zero mass and entropy as required.
Global $AdS_2$ corresponds to the solution $t=cot(-i u/\ell)$. Using eq. (12) we get the mass of this solution to be
$M=-16 \pi G_2/ \ell$ as expected.

\bigskip
\centerline{\bf 4. $AdS_2$ Black Holes and $NAdS_2$ as the Same States of a 1D CFT} 
\medskip

Further evidence for the one to one correspondence between $AdS_2$ black holes and $NAdS_2$ spacetimes comes from the fact that both sets of solutions can be described as states with the same conformal weights of 1D CFTs with the same central charge. This means that for every dilatonic $AdS_2$ black hole there is a corresponding $NAdS_2$ spacetime with exactly the same properties.

It is well--known that, as a result of the AdS/CFT duality[\ADS], $AdS_2$ black holes can be described as states of a 1D boundary CFT. 
The details of the boundary CFT are not known but chiral 2D CFTs[\CHI], conformal QM[\CQM], DLCQ CFTs[\VJY] have been considered in the literature. (In fact, our results indicate that the boundary theory is a 1D CFT described by the Schwarzian action.)
The central charge of the CFT can be computed from the asymptotic symmetry of the $AdS_2$ black hole metric to be
$c=3/2 \pi G_2$[\MET]. The conformal weight of the state that
describes a black hole is fixed by its mass as $L_0=M_{BH} \ell=r_h^2 / {16 \pi G_2 \ell^2}$. The Cardy formula then gives the correct black hole entropy[\MET].

$NAdS_2$ spacetimes, on the other hand, are described by the Schwarzian action in eq. (6). In ref. [\PGB] it was argued that this action is precisely the description of a 1D CFT where the coefficient of the action is $-c/12$. The conformal weight of the CFT state that describes an $NAdS_2$ spacetime is determined by the mass of the solution as $L_0=M \ell$ just like for the black holes.

Any CFT state with $L_0 \not=0$ is actually an ensemble of states at a finite temperature which breaks the conformal symmetry
spontaneously. In addition, when the central charge does not vanish, the CFT is anomalous.
The anomaly is given by the commutators of the Virasoro generators
$$[L_n,L_m]=(n-m)L_{n+m}+{c \over {12}} n(n^2-1) \delta_{n+m,0} \quad, \eqno(14)$$
where the second term proportional to the central charge is the anomaly. From eq. (14) it is easy to see that there is no anomaly for 
$L_0, L_{\pm 1}$.
Thus, the full conformal symmetry is anomalously broken down to the global conformal symmetry or $SL(2,R)$. Due to the spontaneous and anomalous breaking of the conformal symmetry, at low energies, the CFT is described by the PGBs whose action is invariant under the unbroken $SL(2,R)$.

The PGBs of conformal symmetry can be described by the reparametrizations 
which realize the conformal transformations nonlinearly, i.e. $t_E-x=u \to \xi(u)$. Note that $u$ is periodic, $u \sim u+ \beta$.
The spatial direction $x$ may or may not be compact as long as its period is $>>\beta$, i.e. the CFT is at a high temperature.
The energy--momentum tensor for $\xi(u)$ is 
$$T(\xi)=L_0^{\prime} \xi^{\prime 2}+{c \over {12}} Sch(\xi,u) \quad. \eqno(15)$$
From this we can deduce the PGB action[\PGB]
$$I({\xi})=- \int_0^{\beta} du \left ({L_0^{\prime}}\xi^{\prime 2}+ {c \over {12}}Sch(\xi,u) \right) \quad. \eqno(16)$$
Any rescaling of a solution to the equation of motion (see eq. (9)) is also a solution. We can use this freedom to normalize the kinetic term so that the action now becomes
$$I({\xi})=- {c \over {12}} \int_0^{\beta} du \left ({1 \over 2}\xi^{\prime 2}+ Sch(\xi,u) \right) \quad, \eqno(17)$$
which is exactly the action for $NAdS_2$ spacetime given by eq. (8).

There is a subtlety that arises when we want to compare the dimensionful $NAdS_2$ quantities such as $M$ and $T$ with the corresponding CFT ones which are dimensionless. This requires normalizing $NAdS_2$ mass and temperature by $M \to M \ell$ and $T \to T \ell$. As a result, we find that the normalized value of the dilaton becomes $\phi_r=1/8 \pi G_2$ and 
therefore $c= 12 \phi_r=3/2 \pi G_2$. This agrees with the central charge of the boundary CFT obtained from the asymptotic symmetry of the $AdS_2$  metric. In addition, the conformal weight is $L_0=M \ell$ which again agrees with that of $AdS_2$ black holes. We thus conclude that both $AdS_2$ black holes and $NAdS_2$ spacetimes are states with identical $L_0$'s of 1D CFTs with the same $c$. This does not necessarily mean that the CFTs are the same but it is enough to establish the one to one correspondence between dilatonic $AdS_2$ black holes and $NAdS_2$ spacetimes.
On the other hand, if the CFTs are the same, our results provide strong evidence for the duality between $AdS_2$ and the 1D boundary CFT described by the Schwarzian action..

\bigskip
\centerline{\bf 5. The Schwarzian Theory and $AdS_2$ Black Holes}
\medskip

We found that, as expected from the AdS/CFT duality, the solutions to the Schwarzian action, namely $NAdS_2$ spacetimes are in one to one correspondence with dilatonic $AdS_2$ black holes. 
However, in 2D, the bulk and boundary solutions are not only dual to each other but are actually the same in the following sense.
As we show below, a field redefinition and identification of the boundary time with the bulk radial coordinate transforms the equation of motion derived from the Schwarzian action into the time independent part of the 2D Einstein equation with a negative cosmological constant. As a result, the function that fixes the location of the deformed boundary in
$NAdS_2$ spacetimes becomes exactly the conformal factor of $AdS_2$ black hole metric.

We begin by defining a new field $e^{\rho}=t^{\prime}$. The deformed boundary of $AdS_2$ is located $z(u)=\epsilon t^{\prime}(u)$ so $e^{\rho}$ is basically the location of the $AdS_2$ boundary as a function of boundary time.
In terms of $\rho$, the Schwarzian action in eq. (6) becomes[\VER]
$$I({\rho})= \int du \left({{\phi_r} \over 2} \rho^{\prime 2}-\lambda e^{\rho}+\lambda t^{\prime}\right) \quad, \eqno(18)$$
where $\lambda$ is a Lagrange multiplier which enforces the transformation between
$\rho$ and $t$ as a constraint. The equation of motion for $t$ simply gives a constant $\lambda$.
The equation of motion for $\rho$ is
$$\phi_r \rho^{\prime \prime}+ \lambda e^{\rho} =0 \quad, \eqno(19)$$
which is exactly the time independent part of the (Euclidean) 2D Einstein equation with negative cosmological constant, $\Lambda<0$, written in the conformal gauge 
$$R=-e^{-\rho(x_1)}(\partial_0^2+\partial_1^2) \rho(x_1)=  \Lambda \quad. \eqno(20)$$
Thus, we see that the field $e^{\rho}$ is exactly the conformal factor of the 2D metric.

We now realize that we need to switch to Lorentzian boundary and bulk times since the $NAdS_2$ solution in Euclidean boundary time does not reproduce the $AdS_2$ black hole metric. In Lorentzian signature, $u=iu_L$, $t=it_L$, so the $NAdS_2$ solution becomes $it_L=cot(i \pi T u_L)$ or $t_L=-coth(\pi T u_L)$. Using Lorentzian boundary time $u_L$ switches the sign of the first term in eq. (19). Comparing eqs. (19) in Lorentzian time and (20) we see that the cosmological constant is $\Lambda=-\lambda/\phi_r=8 \pi G_2 \lambda$ where $\lambda>0$ for in order to get the $AdS_2$ spacetime. $\lambda$ is a free parameter but if we want to identify $NAdS_2$ solutions with dilatonic $AdS_2$ black holes of the same temperature, we need to choose $8 \pi G_2 \lambda=2/\ell^2$. This guarantees that the cosmological constants in eqs. (1) and (19) match.
Now for eqs. (19) and (20) to match we need to identify the Lorentzian boundary time with the radial bulk coordinate, i.e. 
$u_L=-iu=x_1$. This establishes that every solution to the 2D Einstein equation is also a solution of the Schwarzian theory. More precisely, the equation that describes the deformations of the $AdS_2$ boundary is exactly the time independent part of 2D Einstein equation with $\Lambda<0$. As we will show below, the deformation of the $AdS_2$ boundary fixed by $z(u_L)=t^{\prime}(u_L)$ is exactly the conformal factor, $e^{\rho(x_1)}$, that corresponds to the metric of dilatonic $AdS_2$ black holes.

Consider the thermal $NAdS_2$ solution with $t_L=-\alpha coth(\pi T u)$ where we included a factor $\alpha$ in front which is allowed due to $SL(2)$ invariance of the Schwarzian action. We can always assume that $\alpha=\pi T \ell^2$ where $\Lambda=-2/\ell^2$ and thus
$t_L=-\pi T \ell^2 coth(\pi T u)$. The conformal factor arising from $t_L(u_L)$, which is a solution to eq. (19) by construction, is
given by 
$$e^{\rho}= {{\partial {t_L}} \over {\partial {u_L}}}={{(\pi T \ell)^2} \over {sinh^2(\pi T u_L)}}={{(\pi T \ell)^2} \over {sinh^2(\mu x_1)}}  \quad. \eqno(21)$$
This is exactly the conformal factor for the $AdS_2$ black hole metric given in eq. (3) where we used 
$\pi T_{BH}=\mu$. This establishes the equivalence between dilatonic $AdS_2$ black holes with $M_{BH},T_{BH}$ given by eq. (10) and $NAdS_2$ spacetimes with the same $M$ and $T$. 

Since the location of the $NAdS_2$ boundary and the conformal factor are given by $z= \epsilon t_L^{\prime}$ and $e^{\rho}=t_L^{\prime}$ respectively, we found that they are identical (up to a constant $\epsilon<<1$ and with the identification $u_L=x_1$). We see that the deformations of the $AdS_2$ boundary as a function of Lorentzian boundary time are described by the the same function that gives the conformal factor of the $AdS_2$ black hole metric in terms of the radial bulk dimension. The (near) boundary region of the black hole metric (at $x_1=0$) or very small Lorentzian boundary times correspond to very large deformations of the $AdS_2$ boundary. On the other hand, the black hole horizon at $x_1 \to \infty$ or large Lorentzian boundary times correspond to very small deformations.

Moreover, the dilaton profile $\phi(x_1)=\mu \ell coth(\mu x_1)$ is exactly $|t(u_L)|$ (up to a rescaling by $\ell$ to make dimensions match). The relation between the conformal factor ($t_L^{\prime}$) and the dilaton ($|t_L|$) is a direct result of
the dilaton equation of motion in 2D dilatonic gravity in the conformal gauge, i.e. $(e^{-\rho} \phi^{\prime})^{\prime}=0$. This shows that the conformal factor is the derivative of the dilaton up to a multiplicative constant as above. Thus, it seems that the solution of Schwarzian theory has all the information necessary to reproduce 2D dilatonic black holes.

This result should be contrasted with the holographic description of AdS black holes in higher dimensions. In the two best understood
cases, i.e. BTZ black holes as states of a 2D CFT[\BTZ] on $S^1$ and $AdS_5$ black holes as thermal states of 4D SYM theory on $S^1 \times S^3$[\WIT],the boundary solutions are clearly not the same as the bulk ones. Moreover, there is no clear, direct way to obtain the black hole metrics from the CFT states. For example, the thermal SYM solution on $S^1 \times S^3$ does not become the $AdS_5$ black hole metric after identifying the boundary time direction with the holographic radial direction.
The equivalence of the bulk and boundary solutions seems to be unique to 2D dilatonic gravity. Its crucial property, for our purposes, is the fact that the metric, i.e. the conformal factor, and the boundary solution are determined by a single function of one variable.

The same arguments apply also to the Poincare patch and global $AdS_2$. Consider the $NAdS_2$ solution that corresponds to the Poincare patch, i.e. $t=\ell^2/u$ or $t_L=-\ell^2/u_L$. The conformal factor arising from this is $e^{\rho}=\ell^2/u^2=\ell^2/x_1^2$. which is the correct conformal factor of the Poincare patch. For global $AdS_2$, the $NAdS_2$ solution is 
$t=\ell cot(-i u/\ell)$ or $t_L=-cot(u_L/\ell)$ which leads to $e^{\rho}=1/ sin^2(x_1/\ell)$ which matches the conformal factor of global $AdS_2$.

$AdS_2$ black holes, the Poincare patch and global $AdS_2$ constitute all the time independent solutions of the 2D Einstein equation with a negative cosmological constant. We saw that these are also the solutions of the boundary theory described by the Schwarzian action. However, the most general solutions to eq. (20) are time dependent.
In the conformal gauge, these are given by the conformal factor[\EIN]
$$e^{\rho}={{4 \ell^2 f^{\prime}(x^+) g^{\prime}(x^-)} \over {[f(x^+)- g(x^-)]^2}} \quad, \eqno(22)$$
where $x^{\pm}=x_0 \pm x_1$ and $f(x^+)$ and $g(x^-)$ are two arbitrary functions that describe the conformal transformations for the right and left movers. It is clear that the 1D boundary theory described by the Schwarzian action cannot
have eq. (22) as a solution since the former depends on one parameter ($u$ or $x_1$) whereas the latter depends on two parameters ($x^{\pm}$). However, by the AdS/CFT duality there must be states in the boundary Schwarzian theory that are dual to the solution in eq. (22). In fact, generating an extra bulk dimension by strong interactions is one of the the magical aspects of the AdS/CFT duality.
It would be very interesting to identify the states of the Schwarzian theory that correspond to the solutions in eq. (20). The existence of such states would be further evidence for the duality between $AdS_2$ and the Schwarzian theory.


\bigskip
\centerline{\bf 6. Conclusions and Discussion}
\medskip

In this paper, we claimed that there is a one to one correspondence between $NAdS_2$ spacetimes and dilatonic $AdS_2$ black holes.
First, we showed that the thermodynamics of the two sets of solutions are the same. Second, we showed that both can be described as states with the same conformal weights in 1D CFTs with the same central charge. These two results establish the one to one correspondence between the two sets of solutions which naively seem very different.  Moreover, we showed that 
$NAdS_2$ spacetimes and dilatonic $AdS_2$ black holes are actually the same set of solutions in the following sense.
By a field reparametrization, the Schwarzian equation of motion can be transformed into the time independent part of the 2D Einstein equation with a negative cosmological constant. As a result, the location of the deformed $AdS_2$ boundary in $NAdS_2$ spacetimes
are given by exactly the conformal factor of the dilatonic $AdS_2$ black hole metric.

Of course, in the context of the AdS/CFT duality, our results would not be surprising if the 1D boundary theory dual to $AdS_2$ is described by the Schwarzian action. Our results, which show a one to one correspondence between the bulk $AdS_2$ black holes and the deformations of the boundary given by $NAdS_2$ spacetimes provide evidence for this duality. Clearly more work, such as a comparison of bulk and boundary correlators and of masses of bulk fields with dimensions of boundary operators, is required to establish this duality.
 

$NAdS_2$ spacetimes are crucial for computing near extremal black hole entropy[\MAL]. Our results mean that the same is true for $AdS_2$ black holes. More specifically, the nonextremal part of the entropy of a near extremal black hole is given by $AdS_2$ black holes exactly the same way it is by $NAdS_2$ spacetimes. 
On the other hand, near extremal black hole entropy can also be counted in the near horizon Rindler space in terms of a horizon CFT[\CARL,\HOR]. It seems that these two descriptions are related. In fact, Rindler space and $AdS_2$ black holes are related by a coordinate transformation (up to a conformal factor)[\CON]. Moreover, they can both be described by a 1D CFT
or the Schwarzian action[\RIN]. It seems that a better understanding of black hole entropy requires clarifying these relationships between $AdS_2$ black holes, Rindler space, 1D CFTs and the Schwarzian theory.

We found that the Schwarzian theory is also described by the action in eq. (18) which is basically the Liouville action (giving rise to the Liouville equation in eq. (19))[\LIO]. In ref. [\RIN], it was shown that generic nonextremal black holes can be described by the Schwarzian theory. Our results indicate that they are also described by Liouville theory. It is interesting to note that Liouville theory also arises in the near horizon decription of nonextremal black holes[\SOL,\EDI]. The solutions in both cases seem to be related which is not surprising since these are two different descriptions of Rindler space.

As mentioned above, our results crucially depend on the unique properties of 2D gravity. Thus, we do not expect similar results in higher dimensions. However the $D=3$ case, i.e. the $AdS_3/CFT_2$ duality seems to be a special case. It is 
well--known that BTZ black holes can be described by states of the boundary CFT that lives on a circle[\STR]. This CFT is a sum of two chiral CFTs each of which can be described by the Schwarzian action. Thus, the boundary states that describes BTZ black holes correspond to two copies of the thermal solution to the Schwarzian action[\NEW]. Whether the identification of the boundary CFT as a product of two Schwarzian theories leads to more insight into BTZ black holes remains an open question.


\bigskip
\centerline{\bf Acknowledgments}

I would like to thank the Stanford Institute for Theoretical Physics for hospitality.

\vfill

\refout

\end
\bye